\documentclass[aps,prl,reprint,groupedaddress]{revtex4-1}
\pdfoutput=1
\usepackage{ae,amsmath,amssymb,amsfonts,bbm,amsthm}
\usepackage{braket}
\usepackage{graphicx}
\usepackage{subfigure}
\usepackage{calc}
\usepackage{color}
\usepackage{transparent}
\usepackage{multirow}
\newcommand{\1}{\ensuremath{\mathbbm{1}}}
\begin{document}

\newtheorem*{lem}{Lemma}
\newtheorem{lemma}{Lemma}[section]

\title{3/4-efficient Bell measurement with passive linear optics and unentangled ancillae}

\author{Fabian Ewert}

\email[]{ewertf@uni-mainz.de}
\affiliation{Institute of Physics, Johannes-Gutenberg Universität Mainz, Staudingerweg 7, 55128 Mainz, Germany}

\author{Peter van Loock}

\affiliation{Institute of Physics, Johannes-Gutenberg Universität Mainz, Staudingerweg 7, 55128 Mainz, Germany}

\date{\today}

\begin{abstract}
It is well known that an unambiguous discrimination of the four optically encoded Bell states is possible with a probability of $50\%$ at best, when using static, passive linear optics and arbitrarily many vacuum mode ancillae. By adding unentangled single-photon ancillae, we are able to surpass this limit and reach a success probability of at least $75\%$. We discuss the error robustness of the proposed scheme and a generalization to reach a success probability arbitrarily close to $100\%$.
\end{abstract}

\maketitle

\paragraph{Introduction}
Bell measurements (BMs), projections of two-qubit states on the Bell basis, form important components of many protocols in quantum computation \cite{KLM01} and communication \cite{DLCZ}. Some of the most prominent examples are quantum teleportation \cite{Be93,Go99}, entanglement swapping \cite{Zu93}, and dense coding \cite{Ma96}. In this paper, we consider Bell states in dual-rail encoding, which is the most convenient and common encoding in optical quantum computation, typically realized through two orthogonal polarization modes. It is well known that an unambiguous BM
in this encoding, utilizing fixed arrays of passive, linear optical elements, arbitrarily many vacuum ancilla modes, and photon number resolving detectors (PNRDs), cannot reach a success probability higher than $50\%$ \cite{Ca01}.
While nonlinear optical interactions of cubic \cite{Kim01} or quartic \cite{Ch95} order do, in principle, allow for a complete BM, such schemes are very inefficient in practice. Of course, techniques from linear-optics quantum computation \cite{KLM01} also enable one to achieve near-unit BM efficiencies, but at the expense of complicated entangled ancilla states and feed-forward operations. Much more recently, two new schemes towards more practical and efficient BMs were presented. On the one hand, Grice \cite{Gr11} demonstrated that a 100\%-efficient BM can be approached without feed-forward, but with sufficiently many entangled (Bell- and GHZ-type) ancilla states, which are still fairly expensive and can be generated only probabilistically. On the other hand, active optical elements such as squeezers, 
without feed-forward and without any ancillae, allow for BMs with above-1/2 efficiency \cite{Za13}. In fact, squeezing still transforms the mode operators linearly and it has also become a viable experimental resource, but such squeezing-enhanced BMs have not yet been shown to reach success probabilities greater than $64.3\%$ \cite{Za13}.

We present in this work a scheme that reaches a success probability of $75\%$ without using any one of the experimentally challenging methods mentioned above. The only resources required are 50:50 beam splitters, PNRDs, and unentangled single photons as ancillae \footnote{compared to Grice \cite{Gr11} who uses two extra entangled photons (one extra Bell pair) to reach 75\% BM efficiency, in our scheme we will need four extra unentangled photons to obtain a value of 75\%. The main practical advantage of our scheme over Grice's \cite{Gr11} then becomes manifest when deterministic single-photon sources are employed \cite{Yo13}, as opposed to a probabilistically generated Bell pair. In order to benefit from our approach, such unconditional single-photon sources do not have to produce ideal pure Fock states. For instance, purities of more than 90\% for detector efficiencies greater than 95\% would suffice in principle, as we will show in the second-to-last section of this paper. Of course, four unconditionally prepared ancilla photons can also be turned into one ancilla Bell pair by using the methods of linear-optics quantum computation \cite{KLM01}. However, this transformation is again non-deterministic, depending on the detection of two photons at the output of two non-deterministic, nonlinear sign shift gates. For the case of heralded single photons, our scheme would need a four-photon detection to herald four ancilla photons, whereas the standard linear-optics approach \cite{KLM01}
for a Bell-pair creation would require a six-photon detection.}.

We further discuss a generalization to reach success probabilities close to $100\%$. This extension is an adaption of the scheme by Grice \cite{Gr11} from ancilla states with at most one photon per mode to those with up to two photons and, unfortunately, it also needs entanglement in the added ancillae. Numerical investigations strongly suggest that these states cannot be obtained from unentangled states with passive linear optics, but techniques with an ancillary atom exist \cite{Sh13}. Finally, we investigate the robustness of our scheme to typical experimental errors such as imperfect photon sources and lossy detectors.

\paragraph{3/4-efficient Bell Measurement}
The Bell states in dual-rail encoding are given by
\begin{align}
\ket{\psi^\pm} &= \frac{1}{\sqrt{2}}\left(\ket{1001}\pm\ket{0110}\right),\\
\ket{\phi^\pm} &= \frac{1}{\sqrt{2}}\left(\ket{1010}\pm\ket{0101}\right).
\end{align}
We label the four optical modes from A to D. The simplest way to do a BM that reaches the 1/2-limit for linear optics is to use two beam splitters, whose action on the mode creation operators is defined by
\begin{align}
	\begin{pmatrix}
		a_1^\dagger \\ a_2^\dagger
	\end{pmatrix}
	\rightarrow \frac{1}{\sqrt{2}}
	\begin{pmatrix}
		1 & i \\ i & 1
	\end{pmatrix}
	\begin{pmatrix}
		a_1^\dagger \\ a_2^\dagger
	\end{pmatrix}.
\end{align}
Throughout this paper a beam splitter always refers to this phase-free 50:50 beam splitter. Applying two of these to combine modes A with C and B with D respectively converts the Bell states to the following form:
\begin{align}
\ket{\psi^+} &\rightarrow \frac{i}{\sqrt{2}}\left(\ket{1100}+\ket{0011}\right),\\
\ket{\psi^-} &\rightarrow \frac{1}{\sqrt{2}}\left(\ket{1001}-\ket{0110}\right),\\
\ket{\phi^\pm} &\rightarrow \frac{i}{2}\left(\ket{2000}+\ket{0020}\pm\ket{0200}\pm\ket{0002}\right).
\end{align}
With a PNRD for each of the modes it is now possible to perfectly discriminate $\ket{\psi^+}$ and $\ket{\psi^-}$ from each other and from $\ket{\phi^\pm}$, whereas $\ket{\phi^\pm}$ are indistinguishable from each other. Thus, an overall success probability of $50\%$ can be attained (given an even distribution for the four Bell states).

Our method to obtain higher success probabilities involves the usual linear-optics elements and ancillary photons. To analyze their use it is convenient to split the modes into two pairs [A,B] and [C,D]. From now on these mode-pairs are treated separately but in exactly the same way. Hence the state $\ket{\psi^-}$ can always unambiguously be discriminated from the other Bell states, since it always sends one photon in each mode-pair while the other Bell states send $0$ or $2$. To discriminate the other three Bell states only the mode pair, in which two photons are sent, can be useful. Hence the remaining problem is to discriminate the three states
\begin{align}
\ket{\alpha}:=\ket{11}, \quad \ket{\beta^\pm}:=\frac{1}{\sqrt{2}} \Big(\ket{20}\pm\ket{02}\Big).
\end{align}
To achieve this, we use ancillary photons in the state
\begin{align}
	\ket{\Upsilon_1}=\frac{1}{\sqrt{2}} \Big(\ket{20}+\ket{02}\Big) \quad\Big( = \ket{\beta^+}\Big).
\end{align}
At first glance this is a highly entangled state, whose creation might be hard, but this state can easily be obtained by sending two single photons through a beam splitter (Hong-Ou-Mandel effect \cite{HOM87}). This is the great advantage of our scheme compared to other methods to surpass the 1/2-limit. We only need single photons as ancillae, no nonlinear effects \cite{Ba05}, feed-forward techniques \cite{KLM01}, squeezing \cite{Za13} or entangled ancillae \cite{Gr11}.

\begin{figure}[b]
	\def\svgwidth{0.4\textwidth}
	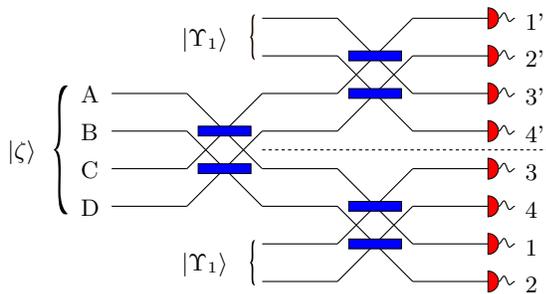
	\caption{Optical setup that identifies an input Bell state $\ket{\zeta}\in\{\ket{\psi^\pm} , \ket{\phi^\pm}\}$ with a success probability of $75\%$. The ancillary state $\ket{\Upsilon_1}$ can easily be obtained form $\ket{11}$ with a beam splitter. It is worth emphasizing that this is a static setup. There is no conditional dynamics between the two halves of the setup so the photon detectors can all be read out simultaneously.	\label{fig:setup}} 
\end{figure}

In the following, the concrete use of the ancillary photons is described. For convenience the modes are now labeled by increasing integers starting with 1 and 2 for the two modes of the relevant pair (see Fig.\ref{fig:setup}). Mixing these modes with the ancillary modes 3 and 4 at two beam splitters (1 with 3 and 2 with 4) leads to
\begin{widetext}
\begin{align}
\ket{\alpha} \ket{\Upsilon_1} & \rightarrow &\! \frac{1}{4\sqrt2}\Big(&-\sqrt3 \ket{3100} + i \ket{2110} - \ket{1120} + i\sqrt3 \ket{0130} -i\sqrt3 \ket{3001} - \ket{2011} - i \ket{1021} -\sqrt3 \ket{0031}\nonumber\\
& & &-\sqrt3 \ket{1300} + i \ket{1201} - \ket{1102} + i\sqrt3 \ket{1003} -i\sqrt3 \ket{0310} - \ket{0211} - i \ket{0112} -\sqrt3 \ket{0013}\Big),\\
\ket{\beta^\pm} \ket{\Upsilon_1} & \rightarrow & \frac18 \Big[&-\sqrt6\Big(\ket{4000}+\ket{0040} \pm\ket{0400}\pm\ket{0004}\Big)-2\Big(\ket{2020}\pm\ket{0202}\Big)\Big]\nonumber\\
& &   + (&1\pm1)\frac18\Big(-\ket{2200} + \ket{2002} +\ket{0220}-\ket{0022}-2\ket{1111}\Big)\nonumber\\
& & +(&1\mp1)\frac{i\sqrt2}{8}\Big(\ket{2101} -\ket{1210}+\ket{1012}-\ket{0121}\Big).
\end{align}
\end{widetext}
Obviously, $\ket{\alpha}$ can be discriminated from $\ket{\beta^\pm}$ unambiguously, since every term that originates from $\ket{\alpha}\ket{\Upsilon_1}$ is unique to $\ket{\alpha}$. It is useful for later to characterize these terms: the total number of photons in modes with an odd label $n_{odd}$ is itself odd. States originating from $\ket{\beta^\pm}\ket{\Upsilon_1}$ on the other hand have even $n_{odd}$.
This is also true in the simple BM without ancillae. The improvement lies in the fact, that there are some unique terms for $\ket{\beta^+}$ and others for $\ket{\beta^-}$. The characterization of these terms is a little more involved. The terms originating from $\ket{\beta^\pm}$ are of two types: either the photons are equally distributed on even and odd modes, or they are all in modes with the same parity. In the latter case no information about the original state can be obtained. But if there are two photons in odd and even modes each, $\ket{\beta^\pm}$ can be discriminated by $n_{[1,2]}$ (the total number of photons in modes 1 and 2), which is even for $\ket{\beta^+} \ket{\Upsilon_1}$ and odd for $\ket{\beta^-} \ket{\Upsilon_1}$.

Adding up the squares of the amplitudes of the unique terms shows, that for each $\ket{\beta^+}$ and $\ket{\beta^-}$ the probability of measuring a unique constellation of photons in the four modes is $50\%$.

If the described optical setup was applied to only one of the original mode pairs (e.g. [A,B]) the success probabilities for $\ket{\phi^\pm}$ were only $25\%$ each, since the setup would only be needed half of the time. It is therefore crucial to use a second pair of ancillary photons on the other side (pair [C,D]). The complete optical setup is shown in Figure \ref{fig:setup}.

The overall success probability of this BM can easily be calculated to be
\begin{align}
	P_{succ}=\frac14 (1 + 1 + 0.5 + 0.5) = \frac34.
\end{align}

\paragraph{Near-unit efficient Bell Measurement}
In this section the presented approach is generalized by adding more ancillary photons to reach higher success probabilities. This section is similar \footnote{Important differences of our generalized scheme from Grice's include: our setup is divided into to halves, where each half gives a new unambiguous-state discrimination problem for three 2-mode states, and this problem is addressed with ancillae having 0 or 2 photons in each mode.} to Grice's general scheme \cite{Gr11} and we shall use a similar notation here.

As before only one mode-pair, e.g [A,B], is considered. The optical setup is defined recursively. Given the setup on the modes $1,...,2^N$ the new setup on the modes $1,...,2^{N+1}$ is constructed as follows: the old setup (without the detectors) is applied to the modes $1,...,2^N$ and an identical copy is also applied to the modes $2^N\!+\!1,...,2^{N+1}$ in which the new ancillary state $\ket{\Upsilon_N}$ is stored. Finally the modes are pairwise mixed: $1$ with $2^N\!+\!1$, $2$ with $2^N\!+\!2$ and so on \footnote{For an illustration of $N=1 \rightarrow N=2$, see the supplementary material.}.

Stated in terms of the $2^{N+1}$-dimensional vector of mode creation operators, we have $\vec{a}^\dagger \rightarrow \mathbf{S}_N \vec{a}^\dagger$ with the matrix $\mathbf{S}_N$ given by the recursive relation
\begin{align}
\mathbf S_N = \frac{1}{\sqrt{2}}\left(\begin{array}{cc}
\mathbf S_{N-1}& i \mathbf S_{N-1}\\
i \mathbf S_{N-1}& \mathbf S_{N-1}
\end{array}\right),
\end{align}
and $\mathbf{S}_0 = \1_{2\times 2}$. The used input state is $\ket{\xi}\ket{\Upsilon_1}\ldots\ket{\Upsilon_{N}}$, a product state of the unknown state $\ket{\xi} \in \{\ket{\alpha}, \ket{\beta^\pm}\}$ and $N$ ancillary states given by
\begin{align}
\ket{\Upsilon_j} :=& \frac{1}{\sqrt{2}\ 2^{2^{j-2}}} \left[\prod_{\substack{k=2^j+1\\ k\text{ odd}}}^{2^{j+1}} (a_k^\dagger)^2 + \prod_{\substack{k=2^j+1\\ k\text{ even}}}^{2^{j+1}} (a_k^\dagger)^2 \right] \ket{\underline{0}}\label{eq:anc}\\=& \frac{1}{\sqrt{2}}\left[\ket{2,0,2,0,\ldots,2,0}+\ket{0,2,0,2,\ldots,0,2}\right], \nonumber
\end{align}
where $\ket{\underline{0}} = \ket{0}\ket{0}...\ket{0}$ denotes the multi-mode vacuum.

Just as in the previous section $\ket{\alpha}$ can be distinguished from $\ket{\beta^\pm}$ using $n_{odd}$: Since each $\ket{\Upsilon_j}$ adds an even number of photons to every mode (0 or 2) and since even and odd modes are not mixed (see definition of $\mathbf S_N$), the parity of $n_{odd}$ does not change from one setup to the next. Thus $n_{odd}$ is odd for $\ket{\alpha}$ and even for $\ket{\beta^\pm}$ for every $N$.

The advantage of the additional ancillary photons is that they allow to reduce the degeneracy of $\ket{\beta^+}$ and $\ket{\beta^-}$ by half in each step. To see this, express the input states for $\ket{\beta^\pm}$ as
\begin{align}
&\ket{\beta^\pm}\ket{\Upsilon_1}\ldots\ket{\Upsilon_{N}}\label{eq:oddevensort}\\
&\quad=\ket{\Xi_{N}^\pm}+\ket{\Xi_{N-1}^\pm}\ket{\Upsilon_{N}}+\ket{\Xi_{N-2}^\pm}\ket{\Upsilon_{N-1}}\ket{\Upsilon_{N}}+\ldots\nonumber\\
&\qquad +\ket{\Xi_{1}^\pm}\ket{\Upsilon_{2}}\ket{\Upsilon_3}\ldots\ket{\Upsilon_{N}}+\ket{\Gamma_N^\pm}, \nonumber
\end{align}
with
\begin{align}
&\ket{\Xi_j^\pm} := \left(\frac{1}{\sqrt{2}}\right)^{j+1} \frac{1}{2^{2^{j-1}}}\times\\
&\times\! \left[\prod_{\substack{k=1\\ k\text{ odd}}}^{2^{j}} (a_k^\dagger)^2 \prod_{\substack{k=2^{j}+1\\ k\text{ even}}}^{2^{j+1}} (a_k^\dagger)^2 \pm \prod_{\substack{k=1\\ k\text{ even}}}^{2^{j}} (a_k^\dagger)^2 \prod_{\substack{k=2^{j}+1\\ k\text{ odd}}}^{2^{j+1}} (a_k^\dagger)^2 \right]\! \ket{\underline{0}},\nonumber
\end{align}
and
\begin{align}
\ket{\Gamma_N^\pm} :=  \left(\frac{1}{\sqrt{2}}\right)^{N+1} \frac{1}{2^{2^{N-1}}}\left[\prod_{\substack{k=1\\ k\text{ odd}}}^{2^{N+1}} (a_k^\dagger)^2 \pm \prod_{\substack{k=1\\ k\text{ even}}}^{2^{N+1}} (a_k^\dagger)^2 \right] \ket{\underline{0}}.
\end{align}
In Eq.\eqref{eq:oddevensort} the terms are sorted by $n_{odd}-n_{even}$: Every $\ket{\Xi_j^\pm}$ leads to an equal number of photons in odd and even modes, but every $\ket{\Upsilon_j}$ adds either $2^{j}$ photons to the odd modes or to the even modes. So the term starting with $\ket{\Xi_j^\pm}$ leads to $n_{odd}-n_{even} = \pm 2^{N} \pm 2^{N-1} \ldots \pm 2^{j+1}$ (and 0 for $j=N$). Further the term $\ket{\Gamma_N^\pm}$ leads to $n_{odd}-n_{even} = \pm 2^{N+1}$.

Hence, all terms in Eq.\eqref{eq:oddevensort} can be distinguished by $n_{odd}-n_{even}$. This can be used to discriminate between $\ket{\beta^+}$ and $\ket{\beta^-}$ since for every term except $\ket{\Gamma_N^\pm}$ the $+$case can be distinguished from the $-$case. To see this consider $\ket{\Xi_N^\pm}$. Since every input operator leads to a linear combination of output operators, each term which derives from the first summand of $\ket{\Xi_N^\pm}$ has the form
\begin{align}
&_{odd,even}\widehat{\Xi}_{\{k1,\ldots,k2^{N+1}\}}\label{eq:oexi}\\
&\quad=\left(S_{[1,k1]}a_{k1}^\dagger S_{[1,k2]}a_{k2}^\dagger S_{[3,k3]}a_{k3}^\dagger \ldots S_{[2^N-1,k2^N]}a_{k2^N}^\dagger  \right)\nonumber\\
&\quad \times \left(S_{[2^N+2,k(2^N+1)]}a_{k(2^N+1)}^\dagger S_{[2^N+2,k(2^N+2)]}a_{k(2^N+2)}^\dagger \right.\nonumber\\&\qquad\quad S_{[2^N+4,k(2^N+3)]}a_{k(2^N+3)}^\dagger \left.\ldots S_{[2^{N+1},k2^{N+1}]}a_{k2^{N+1}}^\dagger  \right), \nonumber
\end{align}
where $S_{[p,q]}$ are the matrix elements of $\mathbf{S}_N$ and where each $kj$ is an integer from $\{1,2,\ldots,2^{N+1}\}$ ("describing in which detector the photon $j$ ended up"). For each term of the form \eqref{eq:oexi} there exists a corresponding term derived from the second summand of $\ket{\Xi_N^\pm}$ with the same set of output operator labels $\{k1,\ldots,k2^{N+1}\}$:
\begin{align}
&_{even,odd}\widehat{\Xi}_{\{k1,\ldots,k2^{N+1}\}}\label{eq:eoxi}\\
&\quad=\left(S_{[2,k(2^N+1)]}a_{k(2^N+1)}^\dagger S_{[2,k(2^N+2)]}a_{k(2^N+2)}^\dagger \right.\nonumber\\& \qquad\quad S_{[4,k(2^N+3)]}a_{k(2N+3)}^\dagger \left.\ldots S_{[2^N,k2^{N+1}]}a_{k2^{N+1}}^\dagger  \right)\nonumber\\
&\qquad\times \left(S_{[2^N+1,k1]}a_{k1}^\dagger S_{[2^N+1,k2]}a_{k2}^\dagger \right.\nonumber\\& \qquad\quad \left. S_{[2^N+3,k3]}a_{k3}^\dagger \ldots S_{[2^{N+1}-1,k2^{N}]}a_{k2^{N}}^\dagger  \right).\nonumber
\end{align}
Combining the terms and using the special form of $\mathbf{S}_N$, particularly
\begin{align}
S_{[j+2^N,kx]}=\left\{\begin{array}{lll}
i S_{[j,kx]}&\text{if}&kx\in\{1,\ldots,2^N\}\\-i S_{[j,kx]}&\text{if}&kx\in\{2^N+1,\ldots,2^{N+1}\}
\end{array}\right.
\end{align}
leads to
\begin{align}
&_{odd,even}\widehat{\Xi}_{\{k1,\ldots,k2^{N+1}\}} \pm _{even,odd}\widehat{\Xi}_{\{k1,\ldots,k2^{N+1}\}}\label{eq:discrim}\\
=&i^{2^N}\left( S_{[1,k1]} S_{[1,k2]}\ S_{[3,k3]} \ldots\ S_{[2^N-1,k2^N]}  \right)\nonumber\\
&\times\left(S_{[2,k(2^N+1)]} S_{[2,k(2^N+2)]} S_{[4,k(2^N+3)]}\ldots\ S_{[2^N,k2^{N+1}]}  \right)\nonumber\\
&\times\left[(-1)^{\tilde{n}_{even}} \pm (-1)^{\tilde{n}_{odd}}\right] \left(a_{k1}^\dagger \ldots a_{k2^{N+1}}^\dagger\right), \nonumber
\end{align}
where $\tilde{n}_{even}$ and $\tilde{n}_{odd}$ are the number of photons in even (odd) modes in the subset $\{2^N+1,\ldots,2^{N+1}\}$ of the outputs. It is clear, that a term of the form \eqref{eq:discrim} appears only for one of the input states $\ket{\beta^\pm}$. More specifically, the parity of $n_{\{1,\ldots,2^N\}}=2^{N+1}-(\tilde{n}_{even}+\tilde{n}_{odd})$ discriminates the two states: even for $\ket{\beta^+}$ and odd for $\ket{\beta^-}$.

To discriminate $+$ from $-$ for the other terms starting with a $\ket{\Xi_j^\pm}$ it is needed, that a discrimination from a smaller setup $(j<N)$ carries over to the current setup. To this end let $A_m^{(p,M)}=\{m2^p+1,\ldots,(m+1)2^p\}$ the subset of the $2^M$ output ports, with $0\leq p\leq M-1$ and $0\leq m\leq 2^{M-p}-1$. Further let $A^{(p,M)}=A_0^{(p,M)}\cup A_2^{(p,M)} \cup \ldots\cup A_{2^{M-p}-2}^{(p,M)}$ and $n^{(p,M)}$ the number of photons in the detector set $A^{(p,M)}$.

\begin{lem} If the $2^M$-photon input state $\ket{\Theta_M}$ always leads to an odd (even) value of $n^{(p,M)}$ in the $2^M$-photon setup, then the input state $\ket{\Theta_M}\ket{\Upsilon_{M+1}}$ always leads to an odd (even) value of $n^{(p,M+1)}$ in the $2^{M+1}$-photon setup. \footnote{A proof of this Lemma can be found in the supplementary material.}
\end{lem}

With the aid of this Lemma it is now clear, that every term except the last in Eq.\eqref{eq:oddevensort} can be used to discriminate $\ket{\beta^\pm}$: As shown above $\ket{\Xi_j^+}$ can be distinguished from $\ket{\Xi_j^-}$ on the $2^{j+1}$-port system by the parity of $n_{\{1,\ldots,2^j\}}=n^{(j,j+1)}$. $\ket{\Xi_j^+}\ket{\Upsilon_{j+1}}$ can then be distinguished from $\ket{\Xi_j^-}\ket{\Upsilon_{j+1}}$ by the parity of $n^{(j,j+2)}$. Repeating this shows, that $\ket{\Xi_j^\pm}\ket{\Upsilon_{j+1}}\ldots\ket{\Upsilon_N}$ can be distinguished by the parity of $n^{(j,N+1)}$. Thus only the terms $\ket{\Gamma_N^\pm}$ in Eq.\eqref{eq:oddevensort} are ambiguous. Their norm is $2^{-N}$ and so the probability to unambiguously identify the states $\ket{\beta^\pm}$ is $1-2^{-N}$ for both, which leads to a total success rate for the Bell-Measurement of
\begin{align}
	P_{succ}^{(N)}=\frac{1+1+2(1-2^{-N})}{4}=1-2^{-N-1},
\end{align}
approaching unity for $N\rightarrow \infty$.

As mentioned before, the ancillary states $\ket{\Upsilon_j}$ are highly entangled and for $j\geq2$ probably cannot be obtained from single-photon states using passive linear optics \footnote{We made various numerical tests with mathematica to obtain the needed ancillae for $N=2$ from single-photon states $\ket{1111}$ and also tried feed-forward techniques using an ancillary photon, but none were successful.}. This seems to imply that $75\%$ poses a boundary to BMs with unentangled ancillae, just like $50\%$ did for BMs with vacuum ancillae. But it turns out that this is not true. With a lengthy but straightforward calculation one can see that a probability of $\frac{25}{32}>\frac{3}{4}$ can be reached. To do this use the setup for $N=2$ but replace $\ket{\Upsilon_2}$ by the state $\ket{\Upsilon_1}\ket{\Upsilon_1}$, which is obtained by sending $\ket{1111}$ through two  beam splitters. Although the gain in success probability surely is not worth the experimental cost, this shows that no conceptual limit has been found yet. \footnote{A more detailed explanation on how to reach $ \frac{25}{32}$ and possibilities to generalize this result are discussed in the supplementary material.}.

\paragraph{Imperfections}
In this section we investigate the influence of errors on the proposed scheme. Although there is a multitude of possible errors, that can occur in quantum optics, we restrict ourselves to two of the main issues: imperfect photon sources and lossy photon detectors. Furthermore we analyze only the 3/4-efficient BM and not the generalized version.

Ideally the photon sources produce the pure state $\ket{1}$. Two of these are sent through a beam splitter to obtain the needed ancilla state $\ket{\Upsilon_1}$. In a more realistic scenario the source will produce a mixed state of the form $\eta_s \ket1\bra1 + (1-\eta_s) \ket0 \bra0$, where $\eta_s$ denotes the probability of the source producing a $\ket{1}$. Combining two of these at a beam splitter leads to
\begin{align}
	& \eta_s^2 \ket{\Upsilon_1}\bra{\Upsilon_1} + (1-\eta_s)^2\ket{00}\bra{00} \nonumber\\&\quad + \eta_s(1-\eta_s) \big(\ket{10}\bra{10} +\ket{01}\bra{01} \big). \label{eq:mixedout}
\end{align}
Since the scheme is not loss resistant, only the first term of this mixture is of use. Thus the success probability needs to be multiplied with a factor of $\eta_s^2$ whenever the ancilla state $\ket{\Upsilon_1}$ is needed.

A lossy photon detector is modeled by a beam splitter with transmittance $\eta_d$ and one empty entry in front of a perfect PNRD. Since only terms without photon loss are of use, the probability of a successful event is multiplied with a factor $\eta_d$ for every photon involved \footnote{For a derivation of this statement see the supplementary.}.

The successful events for the four Bell states can be characterized in the following way:
\begin{itemize}
	\item[$\ket{\psi^+}$] 4 photons in one arm and $n_{odd}$ is odd in this one: $P_{succ}(\ket{\psi^+}) = \eta_s^2\eta_d^4$
	\item[$\ket{\psi^-}$] 3 photons in each arm: $P_{succ}(\ket{\psi^-}) = \eta_s^4\eta_d^6$
	\item[$\ket{\phi^+}$] 4 photons in one arm, $n_{odd}$ is even , $n_{odd}-n_{even}=0$ and $n_{[1,2]}$ is even: $P_{succ}(\ket{\phi^+}) = \frac{1}{2}\eta_s^2\eta_d^4$
	\item[$\ket{\phi^-}$] 4 photons in one arm, $n_{odd}$ is even , $n_{odd}-n_{even}=0$ and $n_{[1,2]}$ is odd:	$P_{succ}(\ket{\phi^-}) = \frac{1}{2}\eta_s^2\eta_d^4$
\end{itemize}
This leads to an overall success probability of
\begin{align}
	P_{succ}(\eta_s,\eta_d) = \frac{1}{2} \eta_s^2 \eta_d^4 + \frac{1}{4} \eta_s^4 \eta_d^6.
\end{align}
It is clear, that success rates higher than $1/2$ can only be obtained with sufficiently good photon sources and detectors. But instead of looking at the bound of an ideal BM without ancillae, i.e. with perfect PNRDs, it makes more sense to compare this result with the success rates of a simple BM as given in the first section, which suffers the same errors. For this simple BM no ancillary photons are needed, but the two photons still need to be detected. In order to be better than the simple BM the experimental parameters of our setup thus need to meet the requirement $\eta_s \eta_d \leq \sqrt{\sqrt{3}-1} \approx 0.86$.

So far only the ancillary state $\ket{\Upsilon_1}$ has been considered useful for the BM. A straightforward analysis, however, shows that also the two-mode vacuum [second term in Eq. \eqref{eq:mixedout}] can be used as an ancilla. Although the vacuum does not help identifying $\ket{\beta^\pm}$, the information about $\ket{\alpha}$ remains intact, unlike the cases with exactly one photon in the ancilla [third term in Eq. \eqref{eq:mixedout}]. Therefore, for heralded photon sources, one could turn the latter cases into vacuum as well through feed-forward, thus beating the simple BM independent of the actual experimental parameters.

\paragraph{Conclusion} We have shown that with the aid of single-photon ancillae the $1/2$-limit for BMs with static, passive linear optics can easily be surpassed and a success probability of more than $3/4$ is possible. This increased success rate has practical relevance, for example, in the creation of cluster states or in quantum repeaters. From a conceptual point of view, it is more important that $\frac{25}{32}$ poses a new maximal value for linear-optics BMs without conditional dynamics, entangled ancillae or active components (squeezing).

\paragraph*{Acknowledgement} We thank Hussain Zaidi for helpful discussions. We also acknowledge support from the BMBF in Germany through QuOReP.

\bibliography{efficient_BM}

\onecolumngrid
\newpage
\section*{Supplementary material}
\renewcommand{\thelemma}{A\arabic{lemma}}
\subsection{No loss of discrimination in bigger setups}
\label{subsec:noloss}
\setcounter{equation}{0}
\renewcommand{\theequation}{S\arabic{equation}}
\begin{lem} If the $2^M$-photon input state $\ket{\Theta_M}$ always leads to an odd (even) value of $n^{(p,M)}$ in the $2^M$-photon setup, then the input state $\ket{\Theta_M}\ket{\Upsilon_{M+1}}$ always leads to an odd (even) value of $n^{(p,M+1)}$ in the $2^{M+1}$-photon setup.
\end{lem}
The proof of this Lemma is almost identical to the one presented by Grice \cite{Gr11} in the appendix. Since there is a small technical mistake which does not affect the validity of the Lemma itself and since we use different ancillae, we give our version of the proof. 
\begin{proof} The proof is based on the structure of $\mathbf{S}_N$, i.e. the structure of the optical setup. Due to the recursive definition of $\mathbf{S}_N$ it follows, that
\begin{align}
	\forall p \in \{1,...,M\}\ \forall j \in A^{(p,M+1)}:\qquad S_{[j+2^p,k]} = \left\{\begin{array}{rl}
	i S_{[j,k]} & \text{if } k \in A^{(p,M+1)}\\ -i S_{[j,k]} & \text{if } k \notin A^{(p,M+1)}
	\end{array} \right.. \label{eq:struct}
\end{align}
(Here Grice used $S_{[j+1,k]}$ which is of course correct for $p=0$, but not in general.) An input creation operator corresponds to a linear combination of output operators. For the $2^M$-photon setup this correspondence is $a_j^\dagger \rightarrow \sum_{k=1}^{2^M} S_{[j,k]} a_k^\dagger$.
The same input operator $a_j^\dagger$ in the $2^{M+1}$-photon setup leads to
\begin{align}
	a_j^\dagger \rightarrow \sum_{k=1}^{2^{M+1}} S_{[j,k]} a_k^\dagger = \sum_{k=1}^{2^M} S_{[j,k]} (a_k^\dagger+i a_{k+2^M}^\dagger),
\end{align}
where the latter follows from the above mentioned structure \eqref{eq:struct} applied to $p=M$. Thus each input operator leads to the same type of output in the $2^{M+1}$-photon setup, as it does in the $2^M$-photon setup, but with $a_k^\dagger$ replaced by $(a_k^\dagger+i a_{k+2^M}^\dagger)$. Since this is consistent with the extension of $A^{(p,M)}$ to $A^{(p,M+1)}$ the parity of $n^{(p,M+1)}$ for the state $\ket{\Theta_M} \otimes \ket{0}^{\otimes 2^M}$ is the same as the parity of $n^{(p,M)}$ for $\ket{\Theta_M}$.

It remains to show, that the ancillary state $\ket{\Upsilon_{M+1}}$ does not change the parity of $n^{(p,M+1)}$ for $p>1$. To see this, it is useful to rewrite the ancilla \eqref{eq:anc} as
\begin{align}
	\ket{\Upsilon_{M+1}} = \frac{1}{\sqrt{2}\ 2^{2^{M-1}}} \left[ \prod_{j\in I_p} \left(a_j^\dagger a_{j+2^p}^\dagger \right)^2 + \prod_{j\in J_p} \left(a_j^\dagger a_{j+2^p}^\dagger \right)^2 \right] \ket{\underline{0}}, \label{eq:anca}
\end{align}
with $I_p = (2 \mathbb{Z} + 1) \cap \{2^M\!+\!1,...,2^{M+1}\} \cap A^{(p,M+1)}$ and $J_p = 2 \mathbb{Z} \cap \{2^M\!+\!1,...,2^{M+1}\} \cap A^{(p,M+1)}$. Consider now a pair of operators
\begin{align}
	a_j^\dagger a_{j+2^p}^\dagger &\rightarrow \sum_{k=1}^{2^{M+1}} S_{[j,k]} a_k^\dagger  \ \sum_{k'=1}^{2^{M+1}} S_{[j+2^p,k']} a_{k'}^\dagger\nonumber\\
	&= \left[\sum_{k\in A^{(p,M+1)}} S_{[j,k]} a_k^\dagger + \sum_{k\notin A^{(p,M+1)}} S_{[j,k]} a_k^\dagger \right] \left[\sum_{k'\in A^{(p,M+1)}} S_{[j+2^p,k']} a_{k'}^\dagger + \sum_{k'\notin A^{(p,M+1)}} S_{[j+2^p,k']} a_{k'}^\dagger \right]\nonumber\\
	&= i \left[\sum_{k\in A^{(p,M+1)}} S_{[j,k]} a_k^\dagger + \sum_{k\notin A^{(p,M+1)}} S_{[j,k]} a_k^\dagger \right] \left[\sum_{k'\in A^{(p,M+1)}} S_{[j,k']} a_{k'}^\dagger - \sum_{k'\notin A^{(p,M+1)}} S_{[j,k']} a_{k'}^\dagger \right]\nonumber\\
	&= i \left[\sum_{k\in A^{(p,M+1)}} S_{[j,k]} a_k^\dagger \right]^2 - i \left[\sum_{k\notin A^{(p,M+1)}} S_{[j,k]} a_k^\dagger \right]^2.\label{eq:oppair}
\end{align}
Here we first split the sums, then apply \eqref{eq:struct} and finally use $(a+b)(a-b)=a^2-b^2$. It is clear from \eqref{eq:oppair} that the two photons either both end up in $A^{(p,M+1)}$, or neither of them does. Thus $\ket{\Upsilon_{M+1}}$ adds an even number of photons to the set $A^{(p,M+1)}$ and the parity of $n^{(p,M+1)}$ is not changed by the addition of $\ket{\Upsilon_{M+1}}$.
\end{proof}

\subsection{Surpassing 75\% with unentangled ancillae}
\begin{figure}[ht]
  \setbox1=\hbox{\def\svgwidth{0.33\textwidth}
	\input{Fig1.pdf_tex}}
  \setbox2=\hbox{\def\svgwidth{0.41\textwidth}
	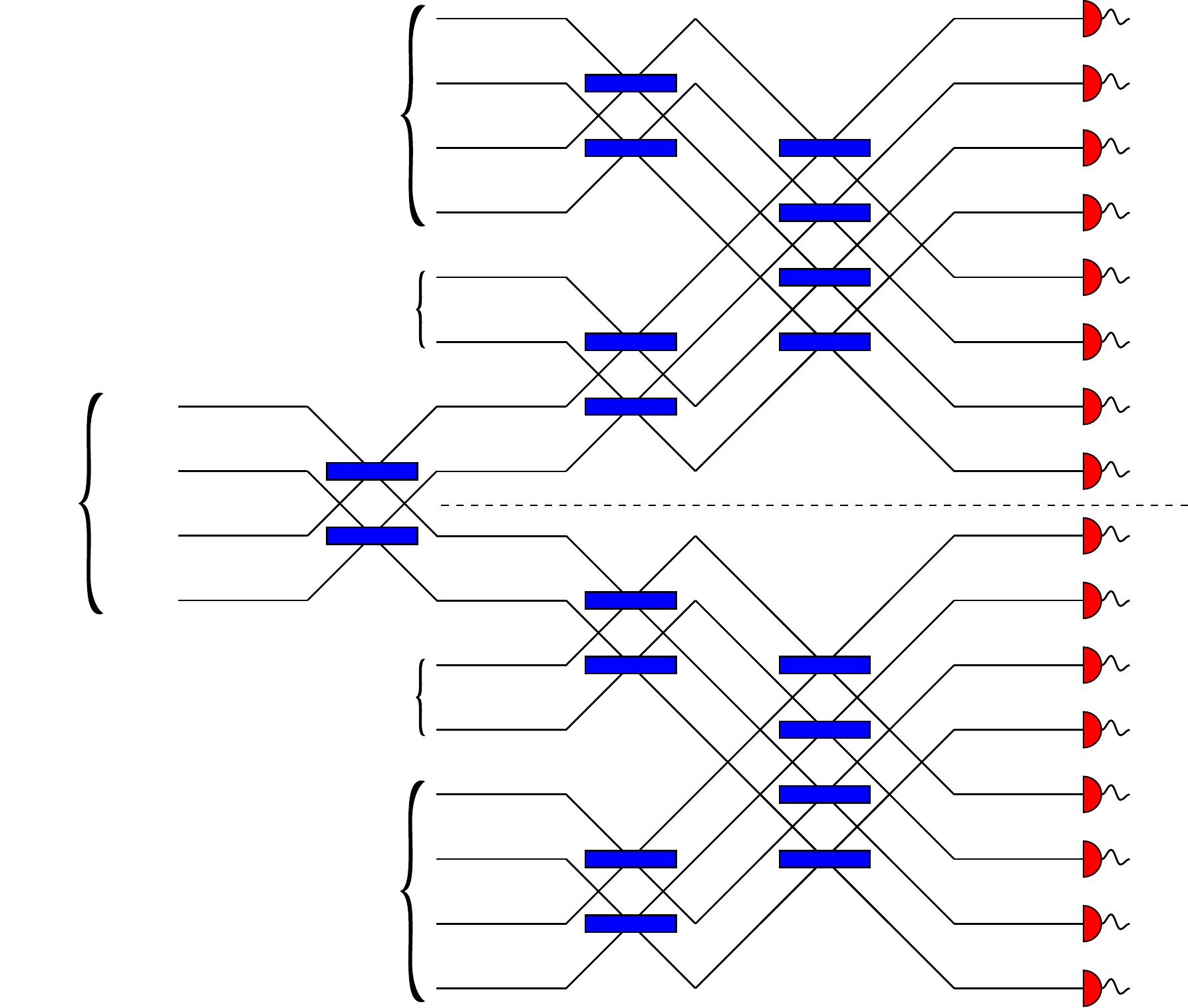}
  {\,} \hfill
  \subfigure[Optical setup for $N=1$	\label{fig:setup3}]{\raisebox{0.5\ht2-0.5\ht1}{\def\svgwidth{0.33\textwidth}
	\input{Fig1.pdf_tex}}} \hfill
  \subfigure[Optical setup for $N=2$	\label{fig:setup2}]{\def\svgwidth{0.4\textwidth}
	\input{Fig2.pdf_tex}} \hfill
  {\,}
  \caption{The above figures show the optical setups for $N=1$ and $N=2$ which lead to success probabilities of $75\%$ and $87.5\%$ respectively. To understand the recursive construction of larger setups it is important to keep in mind that the recursion in the main text considers only one half of the setup.}
\end{figure}

As stated in the main article, it is possible to reach success rates higher than 75\% with single photons as resource. To do so the optical setup for $N=2$ is used (see Fig.\ref{fig:setup2}), but the entangled ancilla $\ket{\Upsilon_2} = 1/\sqrt{2}\ (\ket{2020} + \ket{0202})$ is replaced by $\ket{\Upsilon_1}\ket{\Upsilon_1}$.

Here we need to introduce a slightly more sloppy notation, since in the main article it was clearly defined in which modes the states $\ket{\Upsilon_j}$ are stored. Here the form of the states remains the same, but the modes they are in, is defined by the number of modes occupied by the states to the left. For example in $\ket{\beta^\pm}\ket{\Upsilon_1}\ket{\Xi_1^+}$ the state $\ket{\Xi_1^+}$ is stored in the modes 5,...,8.

In this notation it is clear, that $\ket{\Upsilon_1}\ket{\Upsilon_1}$ can also be written as $1/\sqrt{2}\ket{\Upsilon_2} + \ket{\Xi_1^+}$. Analogously to \eqref{eq:oddevensort} the input state can be split:
\begin{subequations}
\begin{alignat}{2}
	\ket{\beta^\pm}\ket{\Upsilon_1} \ket{\Upsilon_1}\ket{\Upsilon_1} &= \frac{1}{\sqrt{2}}\ket{\beta^\pm}\ket{\Upsilon_1} \ket{\Upsilon_2} &+&\ket{\beta^\pm}\ket{\Upsilon_1}\ket{\Xi_1^+}\nonumber\\
	&= \frac{1}{\sqrt{2}} \ket{\Xi_2^\pm} &+&\ket{\Xi_1^\pm}\ket{\Xi_1^+} \label{eq:75a}\\
	&+ \frac{1}{\sqrt{2}} \ket{\Xi_1^\pm} \ket{\Upsilon_2} &+& \ket{\Gamma_1^\pm}\ket{\Xi_1^+} \label{eq:75b}\\
	&+ \frac{1}{\sqrt{2}} \ket{\Gamma_2^\pm}&&\label{eq:75c}
\end{alignat}\label{eq:75}
\end{subequations}
In \eqref{eq:75} the terms are sorted by $n_{odd}-n_{even}$: 0 for \eqref{eq:75a}, $\pm4$ for \eqref{eq:75b} and $\pm8$ for \eqref{eq:75c}. Thus the three rows can be viewed seperately. Table \ref{tab:par04}, showing which states lead to which parities of $n^{(p,3)}$ can be obtained either by a straightforward calculation, which is rather lengthy and therefore best done numerically, or with the use of the following arguments. The proof of these arguments is left to the interested reader.

\begin{table}[b]
\caption{Parities in the different sets of output operators}
\begin{tabular}{c|cc||c|ccc||c|cc}
	\eqref{eq:75a} & $n^{(1,3)}$ & $n^{(2,3)}$ & \eqref{eq:75b} & $n^{(1,3)}$ & \multicolumn{2}{c||}{$n^{(2,3)}$} & \eqref{eq:75c} & $n^{(1,3)}$ & $n^{(2,3)}$  \\\hline
	$\frac{1}{\sqrt{2}} \ket{\Xi_2^+}$ & even & even & $\ket{\Xi_1^+}\ket{\Gamma_1^+}$ & even & $50/50\ \searrow$ & \multirow{2}{*}{even} & $\frac{1}{\sqrt{2}}\ket{\Gamma_2^+}$ & even & even\\
	$\ket{\Xi_1^+}\ket{\Xi_1^+}$  & even & even & $\ket{\Gamma_1^+}\ket{\Xi_1^+}$ & even & $50/50\ \nearrow$ & & $\frac{1}{\sqrt{2}}\ket{\Gamma_2^-}$ & even & even\\
	$\frac{1}{\sqrt{2}} \ket{\Xi_2^-}$ & even & odd & $\ket{\Xi_1^-}\ket{\Gamma_1^+}$ & odd & $50/50\ \searrow$ & \multirow{2}{*}{$50/50$} &&&\\
	$\ket{\Xi_1^-}\ket{\Xi_1^+}$ & odd & $50/50$ & $\ket{\Gamma_1^-}\ket{\Xi_1^+}$ & even & $50/50\ \nearrow$ &&&&
\end{tabular}
\label{tab:par04}
\end{table}
\setcounter{lemma}{0}
\begin{lemma} The state $\ket{\Xi_M^+}$ leads to even $n^{(p,M+1)}$ for all $1 \leq p \leq M$. Thus it can be used as a replacement for $\ket{\Upsilon_{M+1}}$ in the lemma in section \ref{subsec:noloss}.
\end{lemma}

\begin{lemma} A state of the form $\ket{\Xi_j^+}\ket{\Xi_j^+}$ always leads to even $n^{(j+1,j+2)}$.
\end{lemma}

Analyzing Table \ref{tab:par04} shows, that for $n_{odd}-n_{even}=0$ \eqref{eq:75a} $+$ and $-$ can be discriminated, since for the terms with $-$ at least one discriminator is odd, while for $+$ all are even. For $n_{odd}-n_{even}=\pm4$ \eqref{eq:75b} the terms with $+$ again have  only even discriminators, but for $-$ there are some that have all discriminators even as well (half of the terms originating from $\ket{\Gamma_1^-}\ket{\Xi_1^+}$). Thus the $+$ terms cannot be identified uniquely, and only 3/4 of the terms from $-$ can. As in the original scheme $n_{odd}-n_{even}=\pm8$ doesn't lead to any identifiable terms. Adding up the norm squares of the unique terms gives a success probability of 3/8 for $+$ and of $3/4$ for $-$. Hence the overall success probability is
\begin{align}
\frac{1+1+3/8+3/4}{4} = \frac{25}{32} = 78.125\%.
\end{align}

This result can rather easily be generalized to the following:
\begin{lemma}
When restricted to ancillary states $\ket{\Upsilon_j}$ with $j\leq N$ one can do better than $P_{succ}^{(N)}=1-2^{-N-1}$ by using the $2^{N+2}$-photon setup with the ancillary state $\ket{\Upsilon_1}\ket{\Upsilon_2}...\ket{\Upsilon_{N-1}}\ket{\Upsilon_N} \ket{\Upsilon_N} \ket{\Upsilon_N}$.
This leads to a success probability of $P_{succ}^{(N)}= 1-7/8 \cdot 2^{-N-1}$.
\end{lemma}
Another generalization comes to mind more naturally: replacing all ancillae with products of $\ket{\Upsilon_1}$. For the $N=3$ setup, another four of these ancillae are needed on each side of the setup. The experimental challenge thus grows by a factor of two, but also the theoretical analysis gets more complicated. This is due to the much more involved separation of the possible input states. For example an input state of the form $\ket{\Xi_1^+}\ket{\Upsilon_2}\ket{\Upsilon_2}\ket{\Xi_1^+}$ turns up and can lead to $n_{odd}-n_{even} = 0$ or $\pm8$. Additionally the number of possible combinations of clicks in the detectors grows extremely large. With a combination of analytical and numerical investigation, we found strong hints, that the success probability does NOT increase by adding more $\ket{\Upsilon_1}$. But it should be kept in mind, that this analysis is special to our given optical setup. The $\frac{25}{32}$ therefore do not pose a conceptual limit to passive linear optical BMs with unentangled ancillae.
\subsection{Influence of a lossy photon detector on the efficiency}
As mentioned in the main article a lossy photon detector is modeled by a beam splitter with transparency $\eta_d$ in front of a perfect PNRD. The action of this beam splitter is described by
\begin{align}
	\ket{n} \ket0 \rightarrow \sum_{k=0}^n \sqrt{\binom{n}{k}} \eta_d^{\frac{n-k}{2}} (1-\eta_d)^{\frac{k}{2}} \ket{n-k} \ket{k}.
\end{align}
 Thus a summand $\ket{n_1}\bra{n_2}$ of a mixed state will exit the beam splitter (and then enter the perfect photon detector) as
\begin{align}
	&\text{Tr}_2 \Bigg(\sum_{k_1=0}^{n_1} \sum_{k_2=0}^{n_2} \sqrt{\binom{n_1}{k_1} \binom{n_2}{k_2}} \eta_d ^{\frac{(n_1+n_2-k_1-k_2)}{2}}i^{k_1-k_2} (1-\eta_d)^{\frac{(k_1+k_2)}{2}}\ket{n_1-k_1,k_1}\bra{n_2-k_2,k_2}\Bigg)\nonumber\\
	=& \sum_{k=0}^{\min{(n_1,n_2)}}\sqrt{\binom{n_1}{k} \binom{n_2}{k}} \eta_d ^{\frac{(n_1+n_2-2k)}{2}} (1-\eta_d)^k \ket{n_1-k} \bra{n_2-k}.\label{eq:mixatloss}
\end{align}
Only terms without losses are of use for the presented scheme. These correspond to $k=0$. Since only diagonal entries can be detected by the PNRDs, the probability of a successful event is multiplied with a factor $\eta_d$ for every photon involved.
\subsection{Figures}

\end{document}